\documentclass[aps,prd,eqsecnum,nofootinbib,showpacs,twocolumn]{revtex4-2}

\usepackage{amsmath,amssymb,bm}
\usepackage[dvipdfmx]{graphicx}
\usepackage{color}

\newcommand{\be}{\begin{equation}}
\newcommand{\ee}{\end{equation}}
\newcommand{\bea}{\begin{eqnarray}}
\newcommand{\eea}{\end{eqnarray}}

\newcommand{\eq}[1]{Eq.~(\ref{eq:#1})}
\newcommand{\sect}[1]{Sec.~\ref{sec:#1}}
\newcommand{\appen}[1]{Appendix~\ref{sec:#1}}
\newcommand{\fig}[1]{Fig.~\ref{fig:#1}}

\newcommand{\del}{\partial}

\newcommand{\Tc}{T_c}

\newcommand{\bra}{\langle}
\newcommand{\ket}{\rangle}
\newcommand{\calO}{{\cal O}}

\newcommand{\eg}{{\it e.g.}}

\newcommand{\Nfour}{${\cal N}=4$}

\newcommand{\SC}{superconductor}
\newcommand{\SCs}{superconductors}
\newcommand{\HSC}{holographic superconductor}
\newcommand{\HSCs}{holographic superconductors}

\def\IR{\relax\text{I\kern-.18em R}}

\bmdefine{\bmq}{{\bm{q}}}
\bmdefine{\bmk}{{\bm{k}}}
\bmdefine{\bmx}{{\bm{x}}}
\bmdefine{\bmy}{{\bm{y}}}
\bmdefine{\bmr}{{\bm{r}}}
\bmdefine{\bmnabla}{{\bm{\nabla}}}
\bmdefine{\bmA}{ \bm{A} }
\bmdefine{\bmD}{ \bm{D} }

\bmdefine{\bmPhi}{ \bm{\Phi} }
\bmdefine{\bmPsi}{ \bm{\Psi} }
\bmdefine{\bmcalO}{ \bm{\mathcal{O}} }

\newcommand{\calF}{{\cal F}}

\newcommand{\calL}{{\cal L}}

\newcommand{\calR}{{\cal R}}
\newcommand{\calT}{{\cal T}}

\newcommand{\vecx}{\vec{x}}

\bmdefine{\bmg}{{\bm{g}}}
\bmdefine{\bmR}{{\bm{R}}}

\newcommand{\bwt}{\begin{widetext}}
\newcommand{\ewt}{\end{widetext}}
\newcommand{\bab}{\begin{autobreak}}
\newcommand{\eab}{\end{autobreak}}

\newcommand{\hA}{\Hat{A}}

\newcommand{\calA}{{\cal A}}
\newcommand{\calB}{{\cal B}}
\newcommand{\calG}{{\cal G}}
\newcommand{\calJ}{{\cal J}}
\newcommand{\tilA}{\tilde{A}}

\newcommand{\tilcalA}{\tilde{\calA}}
\newcommand{\tilcalB}{\tilde{\calB}}

\newcommand{\tilcalJ}{\tilde{\calJ}}
\newcommand{\tilcalL}{\tilde{\calL}}

\newcommand{\muo}{\mu_0}
\newcommand{\uo}{r_0}

\begin{document}


\title{Holographic Meissner effect}
\author{Makoto Natsuume}
\email{makoto.natsuume@kek.jp}
\altaffiliation[Also at]{
Department of Particle and Nuclear Physics, 
SOKENDAI (The Graduate University for Advanced Studies), 1-1 Oho, 
Tsukuba, Ibaraki, 305-0801, Japan;
 Department of Physics Engineering, Mie University, 
 Tsu, 514-8507, Japan.}
\affiliation{KEK Theory Center, Institute of Particle and Nuclear Studies, 
High Energy Accelerator Research Organization,
Tsukuba, Ibaraki, 305-0801, Japan}
\author{Takashi Okamura}
\email{tokamura@kwansei.ac.jp}
\affiliation{Department of Physics and Astronomy, Kwansei Gakuin University,
Sanda, Hyogo, 669-1330, Japan}
\date{\today}
\begin{abstract}
The holographic superconductor is the holographic dual of superconductivity, but there is no Meissner effect in the standard holographic superconductor. This is because the boundary Maxwell field is added as an external source and is not dynamical. We show the Meissner effect analytically by imposing the semiclassical Maxwell equation on the AdS boundary. 
Unlike in the Ginzburg-Landau (GL) theory, the extreme Type I limit cannot be reached even in the $e\to\infty$ limit
where $e$ is the $U(1)$ coupling of the boundary Maxwell field. This is due to the bound current which is present even in the pure bulk Maxwell theory. In the bulk 5-dimensional case, the GL parameter and the dual GL theory are obtained analytically for the order parameter of scaling dimension 2.
\end{abstract}
%

\maketitle


\section{Introduction}

The AdS/CFT duality or holography \cite{Maldacena:1997re,Witten:1998qj,Witten:1998zw,Gubser:1998bc} is a useful tool to study strongly-coupled systems (see, \eg, Refs.~\cite{CasalderreySolana:2011us,Natsuume:2014sfa,Ammon:2015wua,Zaanen:2015oix,Hartnoll:2016apf,Baggioli:2019rrs}). 
Let us consider the $(p+2)$-dimensional AdS$_{p+2}$ spacetime and the $(p+1)$-dimensional boundary theory. In the boundary theory, one can add a curved metric and a $U(1)$ Maxwell field, but in most applications, they are not dynamical: one adds them as external sources to the boundary theory. The procedure to promote them to classical dynamical fields has been known \cite{Compere:2008us}. Consider the $(p+1)$-dimensional Einstein equation and the Maxwell equation%
\footnote{We use upper-case Latin indices $M, N, \ldots$ for the $(p+2)$-dimensional bulk spacetime coordinates and use Greek indices $\mu, \nu, \ldots$ for the $(p+1)$-dimensional boundary coordinates. The boundary coordinates are written as  $x^\mu = (t, x^i) =(t, \vecx)=(t,x,y,\cdots)$. 
}:
\begin{subequations}
\label{eq:semiclassical_BC}
\begin{align}
&\calR_{\mu\nu} -\frac{1}{2}\calG_{\mu\nu} \calR = 8\pi G \bra \calT_{\mu\nu} \ket~, \\
&\nabla_\nu \calF^{\mu\nu} = e^2 \bra \calJ^\mu \ket~.
\end{align}
\end{subequations}
All quantities are the $(p+1)$-dimensional ones. The Newton's constant $G$ and the coupling $e$ are the ones for the boundary theory. Here, $\bra \calT_{\mu\nu} \ket$ and $\bra \calJ^\mu \ket$ are expectation values of the boundary energy-momentum tensor and the boundary $U(1)$ current computed by a standard AdS/CFT procedure [\eq{dictionary}]. 
%
%
In other words, one adds the following action to the boundary CFT: 
\begin{align}
S_\text{bdy} = \int d^{p+1}x \sqrt{-\calG}\left( \frac{1}{16\pi G}\calR - \frac{1}{4e^2} \calF_{\mu\nu}\calF^{\mu\nu}\right)~.
%
\end{align}

In standard applications, one imposes the Dirichlet boundary condition on the AdS boundary. For example, for the bulk Maxwell field $A_M$, one imposes  
\begin{align}
\calA_\mu=A_\mu|_{u=0}
%
\end{align}
on the AdS boundary $u\to0$. Instead, we impose the holographic semiclassical equation \eqref{eq:semiclassical_BC} as the boundary condition: we impose the ``mixed" boundary condition.

While the procedure has been known, it has not been studied extensively. One has to consider the bulk equations of motion and the mixed boundary condition simultaneously, and the latter is now a differential equation. In general, it is a difficult task (see, \eg, Ref.~\cite{Ecker:2021cvz} for a recent application). In this paper, we impose the holographic semiclassical equation on the \HSCs\ and show the Meissner effect analytically.

A \HSC\ is typically an Einstein-Maxwell-scalar system \cite{Gubser:2008px,Hartnoll:2008vx,Hartnoll:2008kx}. For $T>\Tc$, the solution is a standard black hole with no scalar, but for $T<\Tc$, the solution becomes unstable and is replaced by a solution with scalar hair. Thus, the scalar corresponds to the order parameter of the phase transition. This is a superconducting transition. For example, the DC conductivity diverges and the London equation holds.

However, in the standard discussion, the boundary Maxwell field is added as a source so is not dynamical. As a result, there is no Meissner effect. The Meissner effect arises from the London equation and the Maxwell equation:
\begin{subequations}
\begin{align}
e^2\calJ_i &= -\frac{1}{\lambda^2} \calA_i~, \\
\del_j \calF^{ij} &= e^2 \calJ^i~.
%
\end{align}
\end{subequations}
Because the latter is absent in the standard \HSC, the Meissner effect does not arise, and a magnetic field can penetrate the \HSC. In a sense, the standard \HSC\ is the ``extreme" Type II \SCs. Or one would regard the system as a superfluid. 

The holographic semiclassical equation for \HSCs\ has been investigated previously \cite{Domenech:2010nf}. The paper studies the issue by constructing a single vortex solution numerically (see, \eg, 
Refs.~\cite{Albash:2009iq,Montull:2009fe,Maeda:2009vf,Keranen:2009re,Dias:2013bwa} for holographic vortices). However, it is desirable to show the Meissner effect analytically.
Our results are summarized as follows:
\begin{enumerate}
\item 
We first consider the case where the condensate is approximately constant and add a magnetic field perturbatively (\sect{small}). 
The boundary current has the supercurrent as well as a contribution from the normal component which exists even in the pure Maxwell theory. The contribution can be interpreted as the bound current, and it changes the magnetic permeability (magnetic constant) from the vacuum value $\muo=e^2$ to $\mu_m$. The magnetic penetration length $\lambda$ and the Ginzburg-Landau (GL) parameter $\kappa$ has a nontrivial $e$-dependence from the magnetic permeability. When $e\ll1$, the result reduces to the standard GL result, but it deviates as one increases $e$. In the $e\to\infty$ limit, $\lambda$ remains finite, and the extreme Type I limit ($\lambda\to0$) cannot be reached.
\item
One often imposes the Neumann boundary condition $\bra\calJ^i\ket=0$ in literature. This corresponds to the $e\to\infty$ limit because $\del_j\calF^{ij}=e^2\bra\calJ^i\ket$. The nontrivial $e\to\infty$ limit explains why one can obtain a Type II \SC\ rather than the extreme Type I \SC\ under the Neumann boundary condition.
\item
In \sect{upper}, we consider the case where the magnetic field is near the upper critical magnetic field $H_{c2}$. We obtain the holographic vortex lattice and show that the magnetic field decreases by the amount $|\psi|^2$, where $\psi$ is the condensate: this also implies the Meissner effect. 
\item
We focus on the $p=2$ case, but the analysis of the $p=3$ case is similar. For $p=3$, an analytic solution is available \cite{Herzog:2010vz,Natsuume:2018yrg}, so one is able to obtain the GL parameter explicitly (\sect{analytic}). Whether the \HSC\ is Type I or Type II depends on $e$ as well as the temperature. Also, we determine the dual GL theory. 
\end{enumerate}

\section{Preliminaries}

We consider the bulk 4-dimensional $s$-wave \HSC:
\begin{subequations}
\begin{align}
S_\text{bulk} &= \int d^4x \sqrt{-g}(R-2\Lambda)+S_\text{m}~, \\
S_\text{m} &= -\frac{1}{g^2} \int d^4x \sqrt{-g} \biggl\{ \frac{1}{4}F_{MN}^2 + |D_M\Psi|^2+m^2 |\Psi|^2 \biggr\}~,
%
\end{align}
\end{subequations}
where 
\begin{subequations}
\begin{align}
F_{MN} &=\del_M A_N -\del_N A_M~, \\ 
D_M &=\nabla_M-iA_M~, \\
\Lambda &=-\frac{3}{L^2}~.
%
\end{align}
\end{subequations}

Below we take the probe limit $g\gg1$ where the backreaction of the matter fields onto the geometry is ignored. Then, the background metric is given by the Schwarzschild-AdS$_4$ (SAdS$_4$) black hole:
\begin{subequations}
\begin{align}
ds_4^2 &= r^2(-fdt^2+dx^2+dy^2)+\frac{dr^2}{r^2f} \\
&=\left(\frac{\uo}{u}\right)^2(-fdt^2+dx^2+dy^2)+\frac{du^2}{u^2f}~, \\
f &= 1-\left(\frac{r_0}{r}\right)^3=1-u^3~, 
%
\end{align}
\end{subequations}
where $u:=r_0/r$.
For simplicity, we set the AdS radius $L=1$ and the horizon radius $r_0=1$. The Hawking temperature is given by $2\pi T =3r_0/(2L^2)$. 
The bulk equations of motion are given by
\begin{subequations}
\begin{align}
0 &= D^2\Psi-m^2\Psi~, \\
0 &= \nabla_NF^{MN} - J^M~,\\
J_M &= -i\{ \Psi^\dag D_M\Psi -\Psi(D_M\Psi)^\dag\} \\
&= 2\Im(\Psi^\dag D_M\Psi)~.
%
\end{align}
\end{subequations}
In the $A_u = 0$ gauge, 
the static bulk equations become
\begin{subequations}
\label{eq:eom_bulk}
\begin{align}
0 =& (-f\del_u^2 - \Delta + 2|\varphi|^2)A_t~, \\
0 =& \{ -\del_u(f\del_u) - \Delta +2|\varphi|^2 \} A_i -2\Im(\varphi^\dag \del_i\varphi) 
\nonumber \\
&+ \del_i(\vec{\del}\cdot\vec{A})~, \\
0 =& \left\{ -\del_u(f\del_u) + V - \frac{A_t^2}{f} - \delta^{ij}D_iD_j \right\} \varphi~, \\
0 =& \del_u(\vec{\del}\cdot\vec{A}) - 2 \Im(\varphi^\dag\del_u\varphi)~,
%
\end{align}
\end{subequations}
where $\Delta:=\del_x^2+\del_y^2$,  $(\vec{\del}\cdot\vec{A}) := \delta^{ij}\del_i A_j$, and
\begin{subequations}
\begin{align}
\Psi &=: u\varphi~,\\
V &:=\frac{m^2+2f-uf'}{u^2}~.
%
\end{align}
\end{subequations}

In the $A_u=0$ gauge, the asymptotic behaviors of matter fields are given by
\begin{subequations}
\label{eq:dictionary}
\begin{align}
A_\mu &\sim \calA_\mu + A_\mu^{(+)} u~, \\
\Psi &\sim \Psi^{(-)} u^{\Delta_-} + \Psi^{(+)} u^{\Delta_+}~, \\
\Delta_\pm &:= \frac{3}{2}\pm\nu~, \quad \nu=\sqrt{\frac{9}{4}+m^2}~.
%
\end{align}
\end{subequations}
$\calA_t=\mu$ is the chemical potential, and $A_t^{(+)}$ represents the charge density $\bra\rho\ket$. 
Similarly, $\calA_i$ is the vector potential, and $A_i^{(+)}$ represents the current density $\bra\calJ_i\ket$. 
$\Psi^{(+)}$ represents the order parameter $\bra\calO\ket$, and $\Psi^{(-)}$ is the external source for the order parameter%
\footnote{For simplicity, we do not consider the ``alternative quantization" where the role of $\Psi^{(-)}$ and $\Psi^{(+)}$ is exchanged \cite{Klebanov:1999tb}.}.
According to the standard AdS/CFT dictionary,
\begin{subequations}
\begin{align}
\bra \calJ_\mu\ket &= \frac{1}{g^2}\left. F_{u\mu} \right|_{u=0}~, \\
\bra \calO \ket &= \frac{1}{g^2}\, 2\nu\Psi^{(+)}~,
%
\end{align}
\end{subequations}
where one needs a standard counterterm action for the scalar field but the counterterm action for the Maxwell field makes no contribution for $p=2$. Although we take the probe limit, we set $g=1$ below for simplicity.

We impose the mixed boundary condition:
\begin{align}
\del_j\calF^{ij}=e^2\bra\calJ^i\ket~.
%
\end{align}
Note that the vacuum magnetic permeability $\muo$ and the vacuum electric permittivity $\epsilon_0$ are given by $\muo=1/\epsilon_0=e^2$. 

\section{Small magnetic field}\label{sec:small}

We would like to know whether a magnetic field can enter the \HSC. Below the critical temperature, a uniform condensate $\varphi_0 =\varphi_0(u)$ is a solution, and we apply a magnetic field there perturbatively. 
%
%
For simplicity, we consider $A_y=A_y(x,u)$ and $B=F_{xy}=\del_x A_y$. We make the Fourier transformation:
\begin{align}
A_y=\int \frac{dq}{(2\pi)} e^{iqx} \tilA_y~.
%
\end{align}
Then, the bulk Maxwell equation becomes
\begin{align}
0 &= \{ -\del_u(f\del_u) +q^2 + 2|\varphi_0|^2 \} \tilA_y~.
\label{eq:eom_Y}
\end{align}

\subsection{Dirichlet boundary condition}

First, let us start with the standard Dirichlet boundary condition. In this case, there should be no Meissner effect, and the magnetic field can enter the \SC\ however small the magnetic field is. One can formally integrate \eq{eom_Y} as
\begin{align}
f \del_u \tilA_y =-\int_u^1 du'\, (q^2+2|\varphi_0|^2) \tilA_y(u')~.
%
\end{align}
Note that the right-hand side has a zero of degree 1 at the horizon $u=1$. Further integrating the equation from the AdS boundary gives
\begin{subequations}
\begin{align}
\tilA_y =& \tilcalA_y - \int_0^u \frac{du'}{f(u')} \int_{u'}^1 du''\, (q^2+2|\varphi_0|^2) \tilA_y(u'') \\
=& \tilcalA_y \left\{ 1- \int_0^u \frac{du'}{f(u')} \int_{u'}^1 du''\, (q^2+2|\varphi_0|^2) +\cdots \right\} \\
=& \tilcalA_y \left\{ 1- q^2\int_0^u \frac{du'}{1+u'+u'^2} \right.
\nonumber \\
&\left. - \int_0^u \frac{du'}{f(u')} \int_{u'}^1 du''\, 2|\varphi_0|^2+\cdots \right\}~.
%
\end{align}
\end{subequations}
We do not evaluate the integral explicitly, but one can obtain the regular solution in principle. The $u'$-integral involves $1/f(u')$, which may give a logarithmic divergence at $u'=1$, but the $u''$-integral has a zero there, so the solution is regular. 

The first term represents the magnetic induction $\tilcalB=iq\tilcalA_y$. From the above result, a $\tilcalB\neq0$ solution exists even when $\varphi_0\neq0$, and the magnetic induction does not decrease. This implies $H_{c1}=0$ for the \HSC\ under the Dirichlet boundary condition.

The remaining terms represent the current. The current is given by
\begin{subequations}
\label{eq:current4}
\begin{align}
\bra\tilcalJ_y\ket &= \del_u \tilA_y|_{u=0} \\
&= \tilcalA_y \left(-q^2 - 2\int_0^1 du\, |\varphi_0|^2+\cdots \right)~.
%
\end{align}
\end{subequations}
This is the London equation with added normal component. 
\begin{enumerate}
\item
The second term represents the supercurrent. Thus, the supercurrent itself exits, but there is no Amp\`{e}re law $\nabla\times \calB=e^2 \calJ$ on the boundary, so there is no Meissner effect. 
\item
The first term exists even in the pure Maxwell theory with $\varphi_0=0$. 
The term is interpreted as the bound current which produces a diamagnetic current as we see below. 
\end{enumerate}
%
%
%
%
%
%
%
%
%
%

\subsection{Holographic semiclassical equation}

We now change the boundary condition and impose the holographic semiclassical equation:
\begin{align}
\del_j \calF^{ij} = e^2 \bra\calJ^i\ket~.
%
\end{align}
However, the holographic semiclassical equation gives
\begin{subequations}
\begin{align}
q^2 \tilcalA_y &= -e^2(q^2+2I)\tilcalA_y \to \tilcalA_y=0~,\\
I &:= \int_0^1 du\,|\varphi_0|^2~.
%
\end{align}
\end{subequations}
Namely, an inhomogeneous magnetic field is not allowed. A similar result holds in the GL theory. 

In order to obtain a nontrivial solution, one must add an external source: 
\begin{align}
\del_j \calF^{ij} = e^2 \bra\calJ^i\ket + e^2 \calJ^i_\text{ext}~.
%
\end{align}
First, let us consider the first term in \eq{current4}, the bound current part. The semiclassical equation is rewritten as 
\begin{subequations}
\begin{align}
q^2 \tilcalA_y &= -e^2q^2\tilcalA_y + e^2 \tilcalJ_y^\text{ext}~, \\
\to 
q^2 \tilcalA_y &= \frac{e^2}{1+e^2}\tilcalJ_y^\text{ext} := \mu_m\tilcalJ_y^\text{ext}~, 
\label{eq:ampere}\\
\to
\mu_m &=  \frac{e^2}{1+e^2}~.
%
\end{align}
\end{subequations}
The left-hand side of \eq{ampere} is $\nabla\times \calB=\nabla(\nabla\cdot \calA)-\nabla^2\calA \to q^2 \tilcalA$, so the equation describes the Amp\`{e}re law $\nabla\times \calB=\mu_m \calJ$. 
Thus, the net effect of the bound current is to shift the magnetic permeability from the vacuum value $\muo=e^2$ to $\mu_m$%
\footnote{The magnetic permeability has been discussed in holographic optics \cite{Amariti:2010jw}, but it differs from our definition. One can show that we essentially use the Landau-Lifshitz definition \cite{LL}.}. 
The magnetic permeability $\mu_m$ and the magnetic susceptibility $\chi_m$ are related by
\begin{subequations}
\begin{align}
\mu_m &=\muo(1+\chi_m)~, \\
\to
\chi_m &=\frac{\mu_m}{\muo}-1 
= -\frac{e^2}{1+e^2} <0~.
%
\end{align}
\end{subequations}
If $\chi_m<0$, a material is diamagnetic.
If $\chi_m>0$, a material is paramagnetic.
In this case, the bound current produces a diamagnetic current.

We now include the supercurrent, 
and the semiclassical equation becomes
\begin{subequations}
\begin{align}
q^2 \tilcalA_y &= \mu_m (-2I\tilcalA_y+\tilcalJ_y^\text{ext})~, \\
\to
\tilcalA_y &\propto \frac{1}{q^2+2\mu_m I}~.
%
\end{align}
\end{subequations}
When $I \neq 0$,
\begin{subequations}
\begin{align}
\calA_y &\propto e^{-x/\lambda}~, \\
\lambda^2 &= \frac{1}{2\mu_m I} = \frac{1+e^2}{2e^2I}~,
%
\end{align}
\end{subequations}
which implies the Meissner effect with magnetic penetration length $\lambda$. 

In the standard GL theory (\appen{GL}),
\begin{align}
\lambda_\text{GL}^2 = \frac{1}{2e^2|\psi|^2}~,
%
\end{align}
and a superconductor is classified by the GL parameter $\kappa_\text{GL}$:
\begin{align}
\kappa_\text{GL}^2 = \frac{\lambda^2}{\xi^2} = \frac{b}{2e^2}~,
%
\end{align}
where $\xi$ is the correlation length of the order parameter. 
In the GL theory, a \SC\ is Type I when $\kappa_\text{GL}^2<1/2$, and
a \SC\ is Type II when $\kappa_\text{GL}^2>1/2$.

\begin{enumerate}
\item
At weak coupling $e \ll 1$, the $e$-dependence coincides with the GL theory. But the result deviates it as one increases $e$ because of the nontrivial magnetic permeability $\mu_m$. 

\item
In particular, in the $e\to\infty$ limit, $\lambda_\text{GL}\to 0$, so $\kappa_\text{GL}\to 0$. Namely, it is the extreme Type I limit and shows the strong Meissner effect. However, our holographic result shows that $\lambda$ remains finite and it implies that the extreme Type I limit cannot be reached%
\footnote{The correlation length $\xi$ is obtained by solving the bulk scalar equation of motion, so it has no $e$-dependence as we see explicitly in \sect{analytic}. }. 
\end{enumerate}

Restoring the horizon radius $\uo$ gives
\begin{subequations}
\begin{align}
\mu_m &= \frac{e^2}{1+e^2/\uo}~,\\
\chi_m &= -\frac{e^2/\uo}{1+e^2/\uo}~,\\
\lambda^2 &= \frac{1}{2\mu_m \uo I} = \frac{1+e^2/\uo}{2e^2I}\frac{1}{\uo}~.
%
\end{align}
\end{subequations}
Note that $e^2$ has scaling dimension 1 in $(2+1)$-dimensions and the horizon radius (or temperature) has scaling dimension 1.

\subsection{Single vortex}

In a Type II \SC, the magnetic field can enter the \SCs\ keeping the superconducting state. The magnetic field enters by forming vortices. 
As one increases the magnetic field, the magnetic field begins to penetrate into the \SC, and vortices appear at the lower critical magnetic field $H_{c1}$. 

Far from the vortex, the condensate is approximately constant and the magnetic field is small. We consider this region and obtain the magnetic field. We take the polar coordinate $d\vecx_2^2=dr^2+r^2d\phi^2$. The $A_\phi=A_\phi(u,r)$ equation becomes
\begin{align}
0= \del_u(f\del_u A_\phi)+r\del_r\left(\frac{1}{r}\del_r A_\phi \right)-2|\varphi_0|^2 A_\phi~.
%
\end{align}
Using the ansatz $A_\phi = U(u)R(r)$, one obtains
\begin{align}
\frac{1}{U} \del_u(f\del_u U)-2|\varphi_0|^2 &= -\frac{r}{R}\del_r \left(\frac{1}{r}\del_r R \right) \nonumber \\
&= -\frac{1}{\lambda^2}~,
%
\end{align}
where $\lambda$ is the separation constant.
The $R$-equation is the standard equation for the vortex, so
\begin{align}
R\propto \sqrt{r} e^{-r/\lambda}, \quad (r\to\infty)~.
%
\end{align}
Thus, $\lambda$ is the magnetic penetration length. The $U$-equation gives
\begin{subequations}
\begin{align}
f\del_uU &= \int_u^1 du'\, (1/\lambda^2-2|\varphi_0|^2)U~, \\
U &= {\cal U}\left\{ 1 + \int_0^u \frac{du'}{f(u')} \int_{u'}^1 du''(1/\lambda^2-2|\varphi_0|^2)+\cdots \right\}~.
%
\end{align}
\end{subequations}
The current is given by 
\begin{subequations}
\begin{align}
\bra \calJ^\phi \ket &= \left. \frac{1}{r^2}\del_uA_\phi \right|_{u=0} \\
&=\frac{R{\cal U}}{r^2} \int_0^1 du\, (1/\lambda^2-2|\varphi_0|^2)+\cdots~.
\label{eq:current_vortex}
\end{align}
\end{subequations}
Imposing the semiclassical equation, one gets
\begin{subequations}
\begin{align}
0&= -\nabla_j\calF^{\phi j}+e^2\bra\calJ^\phi\ket \\
&= \left. \frac{1}{r}\del_r \left(\frac{1}{r}\del_r A_\phi \right) \right|_{u=0} + e^2\bra \calJ^\phi\ket \\
&= \frac{U}{r}\del_r \left(\frac{1}{r}\del_r R \right) + \left. e^2\frac{R}{r^2}\del_u U \right|_{u=0} \\
&\propto 1/\lambda^2+e^2\left(1/\lambda^2-2I \right)~, 
%
\end{align}
\end{subequations}
where $I=\int_0^1 du\,|\varphi_0|^2$. Thus,
\begin{align}
\lambda^2= \frac{1+e^2}{2e^2I}~.
%
\end{align}
Again, $\lambda$ remains finite in the $e\to\infty$ limit. 
This limit corresponds to the Neumann boundary condition $\bra\calJ^i\ket=0$ because $\del_j\calF^{ij}=e^2\bra\calJ^i\ket$. In fact, \eq{current_vortex} gives $\lambda^2=1/(2I)$ under the Neumann boundary condition.

\section{Near upper critical magnetic field}\label{sec:upper}

We discuss a single vortex in previous section. As one increases the magnetic field further, more and more vortices are created, and the vortices form a lattice which is called the vortex lattice. Eventually, the superconducting state is completely broken at the upper critical magnetic field $H_{c2}$. 
Such holographic vortices have been investigated, and we follow Ref.~\cite{Maeda:2009vf} for the construction of the holographic vortex lattice. 

The vortex lattice produces a supercurrent. However, in the standard \HSC, there is no Maxwell equation on the AdS boundary, so the magnetic field can enter the \SC\ not only at vortex cores. We impose the holographic semiclassical equation and show the Meissner effect. 

Near the upper critical magnetic field, the scalar field remains small, and one can expand matter fields as a series in $\epsilon$, where $\epsilon$ is the deviation parameter from the critical point:
\begin{subequations}
\begin{align}
\varphi(\vecx,u) &= \epsilon\varphi^{(1)}+\cdots~, \\
A_t(\vecx,u) &= A_t^{(0)}+\epsilon^2 A_t^{(2)}+\cdots~, \\
A_i(\vecx,u) &= A_i^{(0)}+\epsilon^2 A_i^{(2)}+\cdots~.
%
\end{align}
\end{subequations}

\subsection{Zeroth order}

At zeroth order, \eq{eom_bulk} become
\begin{subequations}
\begin{align}
0 &= \calL_t A_t^{(0)}~, \\
0 &= \calL_V A_i^{(0)} +\del_i(\vec{\del}\cdot\vec{A}^{(0)})~, \\
0 &= \del_u (\vec{\del}\cdot\vec{A}^{(0)})~,
%
\end{align}
\end{subequations}
where
\begin{subequations}
\begin{align}
\calL_t &= -f\del_u^2 - \Delta~, \\
\calL_V &= -\del_u(f\del_u) - \Delta~, 
%
\end{align}
\end{subequations}
so the Maxwell equation gives%
\footnote{It is not clear if one should impose the semiclassical equation for these zero mode solutions. The zero modes satisfy $\del_\nu \calF^{\mu\nu}=0$, so they are not induced by currents. For definiteness, we impose semiclassical equations only on nonzero modes in this paper. }
\begin{subequations}
\begin{align}
A_t^{(0)} &= \mu(1-u)~, \\
A_x^{(0)} &= 0~, \\
A_y^{(0)} &= Hx~.
%
\end{align}
\end{subequations}
\subsection{First order}

At first order,
%
%
the bulk scalar equation becomes
\begin{align}
0 =& \biggl\{ -\del_u(f\del_u) + V - \frac{\mu^2(1-u)^2}{f} 
\nonumber \\
&- \del_x^2 - (\del_y - iHx)^2 \biggr\}  \varphi^{(1)}~.
%
\end{align}
Using the ansatz
\begin{align}
\varphi^{(1)} = e^{iqy}\chi_q(x) \rho(u)~,
%
\end{align}
one obtains
\begin{subequations}
\begin{align}
\left\{ -\del_u(f\del_u)+V-\frac{\mu^2(1-u)^2}{f} \right\} \rho &= -E \rho~, 
\label{eq:eom_rho} \\
\left\{ -\del_x^2+H^2 \left(x-\frac{q}{H}\right)^2 \right\} \chi_q  &= E \chi_q~,
%
\end{align}
\end{subequations}
where $E$ is a separation constant. The regular bounded solution is given by Hermite function $H_n$ as
\begin{align}
\chi_q = e^{-z^2/2}H_n(z)~, \quad
z:= \sqrt{H}\left(x-\frac{q}{H}\right)~,
%
\end{align}
with the eigenvalue
\begin{align}
E=(2n+1)H~.
%
\end{align}
Below we set $n=0$, so 
\begin{align}
\chi_q = \exp\left\{-\frac{H}{2}\left(x-\frac{q}{H}\right)^2 \right\}~.
%
\end{align}
What we obtained is the ``droplet solution," but superpositions of the droplet solution give rise to a vortex lattice solution where a single vortex is arranged periodically. So, consider the general solution
\begin{subequations}
\begin{align}
\varphi^{(1)} &= \rho_0(u)\Sigma(x,y)~,
\label{eq:varphi_0} \\
\Sigma(x,y) &= \int_{-\infty}^\infty dq\, C(q) e^{iqy} \chi_q(x)~.
%
\end{align}
\end{subequations}
Here, $\rho_0$ is the solution of \eq{eom_rho} with $E=H$. 
One can obtain the vortex lattice solution by choosing $C(q)$ appropriately. As discussed in Ref.~\cite{Maeda:2009vf}, the most favorable solution thermodynamically is the triangular lattice for standard \HSCs.

The first order solution \eqref{eq:varphi_0} satisfies
\begin{align}
(\del_y - iA_y^{(0)}) \varphi^{(1)} = i(\del_x - iA_x^{(0)}) \varphi^{(1)}~,
%
\end{align}
so
\begin{subequations}
\begin{align}
2\Im \left[(\varphi^{(1)})^\dag D_x^{(0)}\varphi^{(1)} \right] &= -\del_y|\varphi^{(1)}|^2~,\\
2\Im \left[(\varphi^{(1)})^\dag D_y^{(0)}\varphi^{(1)} \right] &= \del_x|\varphi^{(1)}|^2~,
%
\end{align}
\end{subequations}
or 
\begin{align}
2\Im \left[(\varphi^{(1)})^\dag D_i^{(0)}\varphi^{(1)} \right] = -\epsilon_i^{~j} \del_j|\varphi^{(1)}|^2~,
\label{eq:bulk_current}
\end{align}
where $\epsilon_{xy}=1$. 

\subsection{Second order}

The construction so far has been discussed in Ref.~\cite{Maeda:2009vf}. 
Let us proceed to the second order solution. We now solve the $A_i^{(2)}$ equation and obtain the current $\bra\calJ_i\ket$. We then impose the holographic semiclassical equation $\del_j\calF^{ij}=e^2\bra\calJ^i\ket$ and show the Meissner effect.

The Maxwell equation at second order is given by
\begin{subequations}
\begin{align}
%
0 &= \calL_V A_i^{(2)} + \epsilon_i^{~j} \del_j|\varphi^{(1)}|^2 + \del_i(\vec{\del}\cdot\vec{A}^{(2)})~, \\
0 &= \del_u(\vec{\del}\cdot\vec{A}^{(2)})~,
\label{eq:gauge}
\end{align}
\end{subequations}
where we use \eq{bulk_current}. From \eq{gauge}, $(\vec{\del}\cdot\vec{A}^{(2)})$ does not depend on $u$. Thus, one can choose $\vec{\del}\cdot\vec{A}^{(2)}=0$ by the gauge transformation which does not depend on $u$ so that one can keep the $A_u=0$ gauge.  In momentum space, 
\begin{subequations}
\begin{align}
0 &= \tilcalL_V \tilA_i^{(2)} + i\epsilon_i^{~j} q_j \widetilde{|\varphi^{(1)}|^2}~,\\
\tilcalL_V &= -\del_u(f\del_u)+q^2~.
%
\end{align}
\end{subequations}
Note that $ \widetilde{|\varphi^{(1)}|^2}$ is the Fourier transformation of $|\varphi^{(1)}|^2$ and is not $ |\tilde{\varphi}^{(1)}|^2$.

The second order solution can be constructed exactly, but it can be shown that it is a nonlocal function in the boundary direction \cite{Maeda:2009vf}. 
This is because holographic results correspond to all orders in effective theory expansion. The GL theory takes only the first few terms in the expansion. In fact, at short wavelength, the London equation is replaced by a nonlocal expression known as the Pippard equation. In order to show the Meissner effect, it is enough to take the long-wavelength $q\to0$ limit.

One could use the coordinate $u$, but it is simpler to use the tortoise coordinate $u_*$:
\begin{subequations}
\begin{align}
ds^2 &= \frac{1}{u^2}\left(-fdt^2+\frac{du^2}{f} \right)+\cdots \\
&= \frac{f}{u^2}\left(-dt^2+du_*^2 \right)+\cdots~,\\
du_* &:=\frac{du}{f}~.
%
\end{align}
\end{subequations}
Here, we take $u_*: 0\to \infty$, and $u_*\to \infty$ corresponds to the horizon. Then,
\begin{subequations}
\begin{align}
0 &= \calL^*\tilA_i^{(2)} +g_i~, \\
\calL^* &= -\del_*^2 + q^2 f~, \\
g_i &= i\epsilon_i^{~j} q_j f \widetilde{|\varphi^{(1)}|^2}~.
%
\end{align}
\end{subequations}
Using the bulk Green's function, the solution is formally written as
\begin{subequations}
\begin{align}
\tilA_i^{(2)} &= a_i - \int_0^\infty du_*'\, G(u_*,u_*')g_i(u_*')~, \\
\calL^* G(u_*,u_*') &= \delta(u_*-u_*')~.
%
\end{align}
\end{subequations}
We impose the boundary conditions (1) $G(u_*=0,u_*')=0$, and (2) $\del_*G|_{u_*\to\infty}=0$. 
The first term $a_i$ is the homogeneous solution:
\begin{align}
(-\del_*^2+q^2f)a_i = 0~.
%
\end{align}
We impose the boundary conditions (1) regular at the horizon and (2) $a_i=\tilcalA_i^{(2)}$ at $u=0$. 

One can construct the homogeneous solution by the $q$-expansion:
\begin{align}
a_i = F_0+q^2F_2+\cdots~.
%
\end{align}
The solution which satisfies the boundary conditions is 
\begin{subequations}
\begin{align}
a_i &=\tilcalA_i^{(2)} \left( 1-q^2\int_0^u \frac{du'}{1+u'+u'^2} \right) +O(q^4) \\
&\sim \tilcalA_i^{(2)} (1-q^2 u+ \cdots)~, \quad (u\to 0)~.
%
\end{align}
\end{subequations}

The function $g_i$ is $O(q)$, and it is enough to construct the Green's function at $q=0$:
\begin{align}
-\del_*^2 G = \delta(u_*-u_*')~.
%
\end{align}
Such a Green's function is obtained from two homogeneous solutions. The homogeneous solutions are
\begin{subequations}
\begin{align}
A_b &=u_*~, \\
A_h&=1~, \\
W &:= A_h\del_*A_b-(\del_*A_h)A_b=1~.
%
\end{align}
\end{subequations}
The solution $A_b$ satisfies the boundary condition at the AdS boundary and $A_h$ satisfies the boundary condition at the horizon. Then, the Green's function is given by
\begin{align}
G(u_*,u_*') 
= \left\{
\begin{array}{ll}
A_h(u_*)A_b(u_*') =u_*' & (u_*'<u_*<\infty) 
\nonumber \\
A_h(u_*')A_b(u_*) =u_* & (0<u_*<u_*') 
\nonumber
\end{array}
\right.
%
\end{align}
Thus,
\begin{align}
\tilA_i^{(2)} =& a_i - u_*\int_{u_*}^{\infty} du_*'\, g_i(u_*') - \int_0^{u_*} du_*'\, u_*'g_i(u_*') 
\nonumber \\
&+O(q^3)~.
%
\end{align}


The current is given by
\begin{subequations}
\begin{align}
\bra\tilcalJ_i\ket 
&= \del_u \tilA_i^{(2)} |_{u=0} 
= \del_* \tilA_i^{(2)} |_{u=0} \\
&= \del_*a_i - \int_0^\infty du_*'\, g_i(u_*') \\
&\sim -q^2 \tilcalA_i^{(2)} - i\epsilon_i^{~j} q_j \int_0^1 du\,\widetilde{|\varphi^{(1)}|^2} + O(q^3) 
\label{eq:current} \\
&=  \tilcalJ_i^n +\tilcalJ_i^s~.
%
\end{align}
\end{subequations}
The second term of \eq{current} is the supercurrent. Once again, the supercurrent itself exists even under the Dirichlet boundary condition, but there is no Meissner effect. The first term of \eq{current} exists even for the pure Maxwell theory, and it is interpreted as the bound current. 

%
%

\subsection{Holographic semiclassical equation}

We now impose the semiclassical equation as the boundary condition:
\begin{align}
\del_j\calF^{ij}=e^2\bra\calJ^i\ket~.
%
\end{align}
In momentum space, $\del_j\calF^{ij} =-\Delta\calA_i \to q^2\tilcalA_i$ in the gauge $\del_i\calA^i=0$. Thus, the holographic semiclassical equation becomes
\begin{subequations}
\begin{align}
q^2 \tilcalA_i^{(2)} &= e^2\tilcalJ_i^n + e^2\tilcalJ_i^s~, \\
\to
q^2 \tilcalA_i^{(2)} &= \frac{e^2}{1+e^2} \tilcalJ_i^s:=\mu_m \tilcalJ_i^s~.
%
\end{align}
\end{subequations}
So,
\begin{subequations}
\begin{align}
\tilcalA_i^{(2)} &=  \frac{\mu_m}{q^2} \tilcalJ_i^s \\
&=  - i\frac{\mu_m}{q^2} \epsilon_i^{~j} q_j \int_0^1 du\,\widetilde{|\varphi^{(1)}|^2}~. 
%
\end{align}
\end{subequations}
$\tilcalB$ is then obtained as
\begin{align}
\tilcalB^{(2)} = i\epsilon^{ij} q_i \tilcalA_j^{(2)}
= -\mu_m \int_0^1 du\,\widetilde{|\varphi^{(1)}|^2}~.
%
\end{align}
Going back to the real space,
\begin{align}
\calB^{(2)} 
= -\mu_m \int_0^1 du\,|\varphi^{(1)}|^2~.
%
\end{align}
By adding the zeroth order solution,
\begin{align}
\calB
= H - \epsilon^2\mu_m \int_0^1 du\,|\varphi^{(1)}|^2~.
%
\end{align}
with $H:=\calB_\infty$.

Finally, let us rewrite the result in terms of the operator expectation value $\bra\calO\ket$. Recall
\begin{subequations}
\begin{align}
\Psi &= u\varphi~, \\
\varphi &= \epsilon\varphi^{(1)} +\cdots = \epsilon\rho_0(u)\Sigma+\cdots~, \\
\rho_0 &\sim \rho_0^{(-)} u^{\Delta_{-}-1} + \rho_0^{(+)} u^{\Delta_{+}-1}~,
%
\end{align}
\end{subequations}
so
\begin{align}
\bra\calO\ket =2\nu \Psi^{(+)} =2\nu \epsilon\rho_0^{(+)}\Sigma~.
%
\end{align}
Then, 
\begin{subequations}
\begin{align}
\calB
&= H -  \epsilon^2\mu_m |\Sigma|^2 \int_0^1 du\,|\rho_0|^2 \\
&= H - \frac{\mu_m}{(2\nu)^2} |\bra\calO\ket|^2 \int_0^1 du\,\left|\rho_0/\rho_0^{(+)}\right|^2~,  \\ 
\mu_m &= \frac{e^2}{1+e^2}~.
%
\end{align}
\end{subequations}
Just like in the GL theory \eqref{eq:GL_meissner}, the magnetic induction $\calB$ reduces by the amount $|\bra\calO\ket|^2$ which implies the Meissner effect.

\begin{enumerate}
\item
At weak coupling $e\ll1$, 
\begin{align}
\calB \sim H - e^2|\bra\calO\ket|^2
%
\end{align}
apart from numerical factors, and the $e$-dependence coincides with the GL theory. But the result deviates it as one increases $e$.
\item
In particular, in the $e\to\infty$ limit,
\begin{align}
\calB
\to H -  \frac{1}{(2\nu)^2} |\bra\calO\ket|^2 \int_0^1 du\,\left|\rho_0/\rho_0^{(+)}\right|^2~.
%
\end{align}
Unlike in the GL theory, there is a nontrivial $e\to\infty$ limit. 
\end{enumerate}

We discuss vortex lattices, but the analysis itself does not tell whether the \HSC\ is Type I or Type II.  Using the GL parameter $\kappa$, $H_{c2}$ and thermodynamic critical magnetic field $H_c$ are related by
\begin{align}
H_{c2}=\sqrt{2}\kappa H_c~.
%
\end{align}
When $\kappa^2>1/2$, $H_{c2}>H_c$, and the \SC\ is Type II.
When $\kappa^2<1/2$, $H_{c2}<H_c$, and the \SC\ is Type I.
The existence of a vortex solution itself does not implies that the \SC\ is Type II. Let us lower the magnetic field. For a Type I \SC, the material can ``supercool," namely it can remain the normal state even for $H<H_c$. Then, at $H=H_{c2}$, nucleation occurs, and the vortex lattice forms. 
In order to determine that our \HSC\ is Type I or II, one needs to determine $\kappa$. 

\section{Bulk 5-dimensions}\label{sec:analytic}

The analysis of the bulk 5-dimensional \HSCs\ is similar, but an analytic solution is available for a particular value of $m^2$ \cite{Herzog:2010vz,Natsuume:2018yrg}, and one can obtain $\lambda$, the correlation length $\xi$, and the GL parameter $\kappa$ explicitly. 

\subsection{The GL parameter}

We agin consider a small magnetic field. In this case,
\begin{subequations}
\begin{align}
\tilA_y =& \tilcalA_y \biggl\{ 1 - \int_0^u \frac{u'du'}{f(u')} \int_{u'}^1 du''\, \frac{1}{u''}\left(\frac{q^2}{\uo^2}+2|\varphi_0|^2\right) 
\nonumber \\
&+\cdots \biggr\}~.
%
\end{align}
\end{subequations}
For SAdS$_5$, $\uo=\pi T$. For $p=3$, one must add a counterterm action
\begin{align}
S_\text{CT}= -\int d^4x\, \frac{1}{4g^2}\sqrt{-\gamma}\gamma^{\mu\nu}\gamma^{\rho\sigma}F_{\mu\rho}F_{\nu\sigma} \times \ln(u/\uo)~,
%
\end{align}
where $\gamma_{\mu\nu}$ is the $(p+1)$-dimensional boundary metric. 
The current is then given by
\begin{subequations}
\begin{align}
\bra\tilcalJ_y\ket &= \left. \frac{\uo^2}{u}\del_u \tilA_y - \del_\nu(\sqrt{-\gamma}F^{y\nu}) \times \ln(u/\uo)\right|_{u=0} 
\label{eq:current5_1} \\
&= \tilcalA_y \left(q^2\ln\uo - 2\uo^2\int_0^1 du\, \frac{1}{u}|\varphi_0|^2+\cdots \right)~.
\label{eq:current5_2}
\end{align}
\end{subequations}
Again, the first term of \eq{current5_2} is a bound current, and the second term is the supercurrent. 

We again impose the holographic semiclassical equation with an external source: 
\begin{subequations}
\begin{align}
q^2 \tilcalA_y &=e^2(q^2c_1-2\uo^2I)\tilcalA_y+e^2\tilcalJ_y^\text{ext}~, \\
c_1 &=\ln \uo~, \\
I &= \int_0^1 du\, \frac{1}{u}|\varphi_0|^2~.
%
\end{align}
\end{subequations}
The bound current part is rewritten as 
\begin{subequations}
\begin{align}
(1-c_1e^2) q^2 \tilcalA_y &= e^2 \tilcalJ_y^\text{ext}~, \\
\to 
q^2 \tilcalA_y &= \frac{e^2}{1-c_1e^2}\tilcalJ_y^\text{ext} := \mu_m\tilcalJ_y^\text{ext}~, \\
\to
\mu_m &=  \frac{e^2}{1-c_1e^2} = \frac{e^2}{1-e^2\ln(\pi T)}~.
%
\end{align}
\end{subequations}
The magnetic susceptibility $\chi_m$ is given  by
\begin{align}
\chi_m &= \frac{c_1e^2}{1-c_1e^2} =  \frac{e^2\ln(\pi T)}{1-e^2\ln(\pi T)}~.
%
\end{align}
At $T=0$, $\chi_m<0$ or diamagnetic. As one increases temperature, $\chi_m>0$ or paramagnetic. Then, $\chi_m$ diverges at $e^2\ln(\pi T)=1$, and $\chi_m<0$ at high temperatures. 

Then,
\begin{subequations}
\begin{align}
\tilcalA_y &\propto \frac{1}{q^2+2\mu_m\uo^2 I}~, \\
\calA_y &\propto e^{-x/\lambda}~, \\
\lambda^2 &= \frac{1}{2\mu_mI}\frac{1}{\uo^2}~,
%
\end{align}
\end{subequations}
which implies the Meissner effect.

When $(p,\Delta)=(3,2)$, there exits a simple analytic solution at the critical point \cite{Herzog:2010vz}. The scalar solution is parametrized by a dimensionless parameter $\mu/T$, and $\mu/T=2\pi$ is the critical point. We fix $T$ and vary $\mu$. The solution is
\begin{align}
\varphi_0 = \sqrt{\frac{24}{\uo}(\mu-\mu_c)} \frac{u}{1+u^2}~.
%
\end{align}
See \appen{analytic_sol} for the details. 
The factor $(\mu-\mu_c)^{1/2}$ shows the mean-field behavior with critical exponent $\beta=1/2$. The solution is a special case of a one-parameter family of holographic Lifshitz \SCs\ \cite{Natsuume:2018yrg}. Then, one can evaluate $I$ explicitly :
\begin{subequations}
\begin{align}
I &= \int_0^1 du\, \frac{1}{u}|\varphi_0|^2 = \frac{6}{\uo}(\mu-\mu_c)~, \\
\lambda^2 &= \frac{1}{2\mu_m I}\frac{1}{\uo^2}
\\
&= \frac{1-e^2\ln(\pi T)}{e^2}\frac{1}{12(\mu-\mu_c)}\frac{1}{\pi T}~.
%
\end{align}
\end{subequations}
On the other hand, 
\begin{align}
\xi^2 = \frac{1}{2(\mu-\mu_c)}\frac{1}{\pi T}~,
%
\end{align}
so
\begin{align}
\kappa^2= \frac{\lambda^2}{\xi^2} = \frac{1-e^2\ln(\pi T)}{6e^2}~.
%
\end{align}
The factor $1/6$ was found previously \cite{Natsuume:2018yrg}. 
\begin{figure}[tb]
\centering
\includegraphics{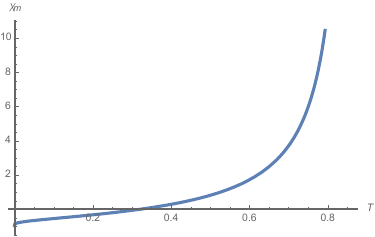}
\includegraphics{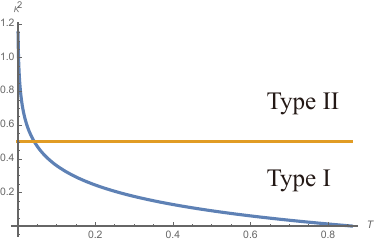}
\vskip2mm
\caption{The magnetic susceptibility $\chi_m$ (top) and the GL parameter $\kappa^2$  (bottom) for $e^2\ln(\pi T)<1$ with $e=1$. 
}
\label{fig:GL}
\end{figure}%
\pagebreak
\begin{enumerate}
\item
Focus on $e^2\ln(\pi T)<1$ where $\mu_m$ and $\lambda^2$ are positive. First, consider a fixed $T$. At weak coupling $e \ll 1$, the $e$-dependence coincides with the GL theory. But the result deviates it as one increases $e$. 

\item
In particular, in the $e\to\infty$ limit, $\lambda_\text{GL}\to 0$, so $\kappa_\text{GL}\to 0$. But in the \HSC,$\lambda$ remains finite and the extreme Type I limit cannot be reached [when $e^2\ln(\pi T)<1$].

\item
In general, whether the \HSC\ is Type I or Type II depends on $T$ as well (\fig{GL}). As $T\to0$, $\kappa\to\infty$, so it is the extreme Type II. As one increases $T$, $\kappa$ decreases. A similar result holds in many superconducting materials including high-$T_c$ materials \cite{Gorkov}. 
\end{enumerate}
The boundary between Type I or II is given by $\kappa^2=1/2$, so 
\begin{subequations}
\begin{align}
T &= \frac{1}{\pi} \exp\left(\frac{1-3e^2}{e^2}\right) \\
&\to \frac{1}{\pi} e^{-3} \sim 0.016~, \quad(e\to \infty)~.
%
\end{align}
\end{subequations}

\subsection{The dual GL theory}

From the results we obtained, one is able to determine the dual GL theory. 
Writing $\psi=\bra\calO\ket$, the GL theory is given by
\begin{align}
F =& \int d^3x \biggl\{ c|D_i\psi|^2+a|\psi|^2+\frac{b}{2}|\psi|^4+\frac{1}{4\mu_m}F_{ij}^2
\nonumber \\
& -(\psi J^\dag+\psi^\dag J) \biggr\}~,
%
\end{align}
where $J$ is the source of the order parameter. From the GL theory, one obtains (see \appen{GL} for the details)
\begin{enumerate}
\item
The spontaneous condensate: $|\psi_0|^2=-a/b$.
\item
The penetration length: $\lambda^2=1/(2c\mu_m|\psi_0|^2)$.
\item 
The correlation length: $\xi^2=-c/a$.
\end{enumerate}
Also, $\kappa^2=b/(2\mu_mc^2)$.

Our holographic results are
\begin{enumerate}
\item
According to the standard AdS/CFT dictionary, $\bra\calO\ket=-\Psi^{(+)}$. The spontaneous condensate is $|\psi|^2=24\epsilon_\mu$, where $\epsilon_\mu := \mu-\mu_c$.
\item
The penetration length: $\lambda^2=2/(\mu_m|\psi|^2)$.
\item 
The correlation length: $\xi^2=1/(2\epsilon_\mu)$.
\end{enumerate}
(For simplicity, we set $\uo=1$). Comparing these results fixes all GL parameters:
\bwt
\begin{align}
F = \int d^3x \left\{ \frac{1}{4}|D_i\psi|^2-\frac{\epsilon_\mu}{2}|\psi|^2+\frac{1}{96}|\psi|^4+\frac{1}{4\mu_m}F_{ij}^2-(\psi J^\dag+\psi^\dag J) \right\}~.
%
\end{align}
\ewt
Just like in the GL theory, this free energy should be regarded as leading terms. For example, we do not include the $O(\psi^6)$ term and higher, and the numerical coefficients are leading ones. Note that the order parameter does not have the canonical normalization. Rather, the normalization is chosen from the GKP-Witten relation. Namely, we fix the normalization of $\psi$ so that the source term is given by $(\psi J^\dag+\psi^\dag J)$ with $J=\Psi^{(-)}$.

\section{Discussion}

\begin{enumerate}
\item
Ref.~\cite{Domenech:2010nf} studies the holographic semiclassical equation by constructing a single vortex solution numerically. However, there is a puzzle in previous analysis. For the $p=2$ \HSC, they consider the Neumann boundary condition $\bra\calJ^i\ket=0$. This is equivalent to the $e\to\infty$ limit because $\del_j\calF^{ij}=e^2\bra\calJ^i\ket$. Then, one expects the extreme Type I \SC, but they obtain a Type II \SC. This was explained in terms of S-duality \cite{Witten:2003ya}. 

Our analysis gives an alternative interpretation. This is because the holographic magnetic penetration length has a nontrivial $e\to\infty$ limit. 
A similar analysis can be done for the $p=3$ case. Thus, the Neumann boundary condition should be possible for the $p=3$ case as well.

\item
In the above analysis, we focus on $e^2\ln(\pi T)<1$, but it is interesting to consider $e^2\ln(\pi T)>1$, where $\mu_m<0$. Using $\epsilon,\mu_m$, one can classify a material as follows:
\begin{enumerate}
\item In the vacuum, $\epsilon, \mu_m>0$, and the speed of light $c$ is given by $c^2=1/(\epsilon \mu_m)$. 
\item For metals, $\epsilon<0,\mu_m>0$, and it implies that the material is not transparent to light. 
\item When $\epsilon<0,\mu_m<0$, the material is transparent to light. Such a material is called a metamaterial and shows the negative refractive index \cite{Amariti:2010jw}.
\item When $\epsilon>0,\mu_m<0$, the material is not transparent to light again. 
\end{enumerate}
The \Nfour\ plasma has 
$\mu_m<0$ for $e^2\ln(\pi T)>1$. Also, one can show that $\epsilon>0$ for the plasma. Thus, the plasma corresponds to (d) and is not transparent to light. It is interesting to study the implications for the quark-gluon plasma. On the other hand, $\lambda^2<0$ in this case, so it implies that the Meissner effect does not occur. Namely, the magnetic field can enter the material. We are unaware if such an effect is discussed in \SC\ literature. 

\item
One limitation of our analysis is that we take the probe limit $g\gg 1$. $g$ appears in the combination $e^2/g^2$, and the $e\to\infty$ limit really means the $e/g\to\infty$ limit. It is certainly interesting to take the backreaction into account, but it is difficult to study the system analytically. 

\item
Finally, we apply the holographic semiclassical equation to the Meissner effect, but it is interesting to explore the other backreaction problems.
\end{enumerate}

\begin{acknowledgments}

This research was supported in part by a Grant-in-Aid for Scientific Research (17K05427) from the Ministry of Education, Culture, Sports, Science and Technology, Japan. 

\end{acknowledgments}

\appendix

\section{Ginzburg-Landau theory}\label{sec:GL}

The GL theory is given by
\begin{subequations}
\begin{align}
F &= \int d^px \left\{ c|D_i\psi|^2+a|\psi|^2+\frac{b}{2}|\psi|^4+\frac{1}{4\mu_m}F_{ij}^2 \right\}~,\\
D_i &:=\del_i-iA_i~,
%
\end{align}
\end{subequations}
where $\mu_m$ is the magnetic permeability. 
In the standard GL theory, $\mu_m=e^2$. Namely, we slightly generalize the GL theory where the material has a magnetization (not due to supercurrent) and there exists a bound current. 
We take
\begin{subequations}
\begin{align}
a&=a_0(T-\Tc)+\cdots~, (a_0>0)~, \\
b&=b_0+\cdots~, (b_0>0)~.
%
\end{align}
\end{subequations}
The equations of motion are given by
\begin{subequations}
\begin{align}
0&=-cD^2\psi+a\psi+b\psi|\psi|^2~,\\
0&=\del_j F^{ij}-\mu_m j^i~,\\
j_i &= - \frac{\delta F_\psi}{\delta A^i} 
\nonumber \\
&=-ic\{\psi^\dag D_i\psi - \psi D_i\psi^\dag\} = 2c\Im[\psi^\dag D_i\psi]~.
%
\end{align}
\end{subequations}
Below $T<\Tc$, a homogeneous spontaneous condensate is a solution:
\begin{align}
|\psi_0|^2 = -\frac{a}{b} \propto \Tc-T~.
%
\end{align}

There are 2 characteristic scales for a superconductor:
\begin{enumerate}
\item
One is the correlation length of the order parameter. This comes from the order parameter mass and is given by
\begin{align}
\xi^2 = \frac{c}{|a|}~.
%
\end{align}

\item
The other is the magnetic penetration length. This comes from the gauge field mass and is given by
\begin{align}
\lambda^2 = \frac{1}{2c\mu_m |\psi_0|^2} = \frac{1}{2c\mu_m}\frac{b}{|a|}~.
%
\end{align}
\end{enumerate}
Then, a superconductor is characterized by a dimensionless parameter $\kappa$, the GL parameter:
\begin{align}
\kappa^2 = \frac{\lambda^2}{\xi^2} = \frac{b}{2\mu_mc^2}~.
%
\end{align}
A superconductor is classified by the value of $\kappa$:
\begin{itemize}
\item Type I: $\kappa^2<1/2$ or $\xi>\sqrt{2}\lambda$.
\item Type II: $\kappa^2>1/2$ or $\xi<\sqrt{2}\lambda$.
\end{itemize}
The factor $1/2$ is determined from the free energy analysis below. In a Type I superconductor, the penetration length is shorter than the correlation length, 
and the magnetic field cannot enter the superconductor. As one increases the magnetic field, eventually superconductivity is broken. In a Type II superconductor, the penetration length is longer than the correlation length, 
and the magnetic field can enter the superconductor keeping the superconducting state. The magnetic field enters by forming vortices. 

Below $T<\Tc$, a homogeneous condensate is a solution. Then, apply a small magnetic field. For simplicity, consider a 2-dimensional \SC\ in the $(x,y)$-plane, and apply a magnetic field in the $z$-direction: $A_y=A_y(x)$ and $B=F_{xy}$. We also add an external source $j_y^\text{ext}$.
The Maxwell equation becomes
\begin{subequations}
\begin{align}
q^2 \tilA_y &= \mu_m( -2c|\psi_0|^2 \tilA_y + j_y^\text{ext})~, \\
\to 
\tilA_y &\propto \frac{1}{q^2+2c\mu_m|\psi_0|^2}~.
%
\end{align}
\end{subequations}
The inverse Fourier transformation gives
\begin{align}
A_y \propto e^{-x/\lambda}~.
%
\end{align}

\subsection{Single vortex}

Far from the vortex, $\psi=\psi_0$ is approximately constant. In the cylindrical coordinate $ds^2=dr^2+r^2d\phi^2+\cdots$, the $A_\phi=A_\phi(r)$ equation becomes 
\begin{align}
0= \frac{1}{r} \left(\frac{1}{r}A_\phi' \right)' - \frac{1}{\lambda^2}\frac{A_\phi}{r^2}~.
%
\end{align}
Then, the solution is given by the modified Bessel function $K_1$:
\begin{align}
A_\phi &= \frac{r}{\lambda} K_1(r/\lambda) \to \sqrt{\frac{\pi r}{2\lambda}}e^{-r/\lambda}~,
%
\end{align}
where we used the asymptotic formula
\begin{align}
K_1(z) \to \sqrt{\frac{\pi}{2z}}e^{-z}~, (z\to\infty)~.
%
\end{align}

\subsection{Critical magnetic field}

The critical magnetic field $H_c$ is defined by the condition that the homogeneous condensate is thermodynamically favorable compared with the normal state. It is convenient to write $\psi=\rho e^{i\theta}$ and use the gauge-invariant variable $\hA_i$:
\begin{align}
\hA_i:= A_i-\del_i\theta~.
%
\end{align}
The free energy becomes
\begin{align}
F = \int d^px \left\{ c(\del_i\rho)^2+(a+c\hA_i^2)\rho^2+\frac{b}{2}\rho^4+\frac{1}{4\mu_m}F_{ij}^2 \right\}~.
%
\end{align}
The equations of motion are given by
\begin{subequations}
\begin{align}
0 &= -c\del_i^2\rho+(a+c\hA_i^2)\rho+b\rho^3~,\\
0 &=\del_jF^{ij} + 2c\mu_m\rho^2\hA_i~.
%
\end{align}
\end{subequations}
The variation of $F$ includes the term
\begin{align}
\delta F= \cdots + \frac{1}{\mu_m}\int dS_i\, F^{ij}\delta \hA_j~.
%
\end{align}
Then, $F$ is appropriate when one fixes $\hA_i$ on the boundary but is not appropriate when one fixes the external magnetic field $F_{ij}= \del_i\hA_j-\del_j\hA_i$. In order to obtain $H_c$, one fixes the external magnetic field, so one should use the Gibbs free energy. We define the Gibbs free energy by
\begin{subequations}
\begin{align}
G &= F - \frac{1}{\mu_m} \int dS_i\, F^{ij} \hA_j \\
&= F - \frac{1}{\mu_m} \int d^px\, \del_i(F^{ij} \hA_j)~.
%
\end{align}
\end{subequations}
Then, the variation becomes
\begin{align}
\delta G = \cdots - \frac{1}{\mu_m} \int dS_i\,  \delta F^{ij}\hA_j~.
%
\end{align}
Using the Maxwell equation, the on-shell Gibbs free energy becomes
\begin{align}
\underline{G} =& \int d^px \biggl\{ c(\del_i \rho)^2+(a+c\hA_i^2)\rho^2+\frac{b}{2}\rho^4 
\nonumber \\
&+2c\hA_i^2\rho^2-\frac{1}{4\mu_m}F_{ij}^2 \biggr\}~.
%
\end{align}

In the superconducting phase, $\rho^2=-a/b$ and $\hA_i=0$ (due to the Meissner effect), so
\begin{align}
\underline{G_s} = -\frac{a^2}{2b}V_p~,
%
\end{align}
where $V_p$ is the $p$-dimensional volume. In the normal phase, $\rho=0$ and $F_{xy}=H$, so
\begin{align}
\underline{G_n} = -\frac{1}{2\mu_m}H^2V_p~.
%
\end{align}
When $\underline{G_s}<\underline{G_n}$, the superconducting phase is favorable, so
\begin{align}
H<H_c = -a\sqrt{\frac{\mu_m}{b}}~.
%
\end{align}
As we see below, $H_{c2}=-a/c$, so
\begin{align}
H_{c2} = \frac{-a}{c} =\sqrt{2}\kappa H_c~.
%
\end{align}
When $\kappa^2<1/2$, $H_{c2}<H_c$, and the \SC\ is Type I.
When $\kappa^2>1/2$, $H_{c2}<H_c$, and the \SC\ is Type II.

\subsection{Upper critical magnetic field}

Near the upper critical magnetic field $H_{c2}$, $\psi$ remains small, and one can expand matter fields as a power series:
\begin{subequations}
\begin{align}
\psi &= \epsilon \psi^{(1)} + \cdots~, \\
A_i &= A_i^{(0)}+\epsilon^2 A_i^{(2)} + \cdots~.
%
\end{align}
\end{subequations}
At zeroth order, the Maxwell equation is
\begin{align}
0= \del_j F_{(0)}^{ij}~,
%
\end{align}
so one has a homogeneous magnetic field
\begin{align}
A_y^{(0)}=Hx~.
%
\end{align}
At first order, the order parameter field obeys
\begin{align}
0= -c(\del_i-iA_i^{(0)})^2\psi^{(1)} +a \psi^{(1)}~.
%
\end{align}
Using the ansatz
\begin{align}
\psi^{(1)} = e^{iqy}\chi_q(x)~,
%
\end{align}
the equation becomes
\begin{align}
c\left\{ -\del_x^2+H^2 \left(x-\frac{q}{H}\right)^2 \right\} \chi_q  = -a \chi_q~.
%
\end{align}
This is the Landau problem, and the solution is given by the Hermite function $H_n$ as
\begin{align}
\chi_q = e^{-z^2/2}H_n(z)~, \quad
z:= \sqrt{H}\left(x-\frac{q}{H}\right)~.
%
\end{align}
The eigenvalue is given by
\begin{align}
E_n=(2n+1)H = \frac{-a}{c}~.
%
\end{align}
$H$ takes the maximum value when $n=0$ or $H_{c2}=-a/c$. The general solution is written as
\begin{align}
\psi^{(1)} &= \int_{-\infty}^\infty dq\, C(q) e^{iqy} \chi_q(x)~.
\label{eq:psi_0}
\end{align}
The first order solution \eqref{eq:psi_0} satisfies
\begin{align}
(\del_y - iA_y^{(0)}) \psi^{(1)} = i(\del_x - iA_x^{(0)}) \psi^{(1)}~,
%
\end{align}
so
\begin{subequations}
\begin{align}
J^{(2)}_x &= 2c\Im \left[(\psi^{(1)})^\dag D_x^{(0)}\psi^{(1)} \right] = -c\del_y|\psi^{(1)}|^2~,\\
J^{(2)}_y &=  c\del_x|\psi^{(1)}|^2~,
%
\end{align}
\end{subequations}
or 
\begin{align}
J^{(2)}_a = 2c\Im \left[(\psi^{(1)})^\dag D_a^{(0)}\psi^{(1)} \right] = -c\epsilon_a^{~b} \del_b |\psi^{(1)}|^2~,
\label{eq:GL_current}
\end{align}
where the Latin indices $a,b$ run though $x$ and $y$, and $\epsilon_{xy}=1$. 
Then, at second order,
\begin{subequations}
\begin{align}
0 &= \del^b F_{ab}^{(2)} - \mu_m J_a^{(2)} \\
&= \epsilon_{ab} \del^b(F_{xy}^{(2)} - c\mu_m |\psi^{(1)}|^2)~.
%
\end{align}
\end{subequations}
One can integrate the equation. Asymptotically, $|\psi^{(1)}|\to0$, so $F_{xy}\to H$. Then,
\begin{align}
F_{xy} &= B = H- c\mu_m |\psi^{(1)}|^2~.
\label{eq:GL_meissner}
\end{align}
Thus, the magnetic induction $B$ reduces by the amount $|\psi^{(1)}|^2$ which implies the Meissner effect. We discuss its holographic counterpart in \sect{upper}.

\section{Analytic solution of \HSC}\label{sec:analytic_sol}

For the SAdS$_{p+2}$ background, the Hawking temperature is $\pi T=(p+1)r_0/4$. In the gauge $A_u = 0$, the static bulk equations become
\begin{subequations}
\begin{align}
0 =& -fu^{p-2}\del_u\left(\frac{1}{u^{p-2}}\del_u A_t \right) - \frac{\Delta}{\uo^2} A_t + 2|\varphi|^2A_t~, \\
0 =& -u^{p-2}\del_u\left(\frac{f}{u^{p-2}}\del_u A_i \right) - \frac{\Delta}{\uo^2} A_i -2\Im(\varphi^\dag D_i\varphi) 
\nonumber \\
&+ \frac{1}{\uo^2}\del_i(\vec{\del}\cdot\vec{A})~, \\
0 =& -u^{p-2}\del_u\left(\frac{f}{u^{p-2}}\del_u \varphi \right) +\left( V - \frac{A_t^2}{\uo^2 f} - \frac{\delta^{ij}D_iD_j}{\uo^2} \right) \varphi~, \\
0 =& \frac{1}{\uo^2}\del_u(\vec{\del}\cdot\vec{A}) - 2\Im(\varphi^\dag\del_u\varphi)~,
%
\end{align}
\end{subequations}
where $\Delta:=\delta^{ij}\del_i\del_j$, $(\vec{\del}\cdot\vec{A}) := \delta^{ij}\del_i A_j$, and 
\begin{subequations}
\begin{align}
\Psi &=: u\varphi~,\\
f &=1-u^{p+1}, \\
V &:=\frac{m^2+pf-uf'}{u^2}~.
%
\end{align}
\end{subequations}
For simplicity, we set $\uo=1$ below.
The asymptotic behaviors of matter fields are given by
\begin{subequations}
\begin{align}
A_\mu &\sim \calA_\mu + A_\mu^{(+)} u^{p-1}~, \\
\Psi &\sim \Psi^{(-)} u^{\Delta_-} + \Psi^{(+)} u^{\Delta_+}~, \\
\Delta_\pm &:= \frac{p+1}{2}\pm\nu~, \quad \nu=\sqrt{\frac{(p+1)^2}{4}+m^2}~.
%
\end{align}
\end{subequations}
When the  the Breitenlohner-Freedman (BF) bound \cite{Breitenlohner:1982bm} is saturated or
\begin{align}
m_\text{BF}^2=-\frac{(p+1)^2}{4}~,
%
\end{align}
the asymptotic behavior is replaced by
\begin{align}
\Psi &\sim \Psi^{(-)} u^{\Delta}\ln u + \Psi^{(+)} u^{\Delta}~, \quad \Delta:=\frac{p+1}{2}~.
%
\end{align}

At high temperature, the equations of motion admit a solution
\begin{align}
A_t = \mu(1-u^{p-1})~, \quad
A_i = 0~, \quad
\Psi =0~. 
\label{eq:sol_high}
\end{align}
But the $\Psi=0$ solution becomes unstable at the critical point and is replaced by a $\Psi\neq0$ solution.

When $(p,\Delta)=(3,2)$, there exits a simple analytic solution at the critical point \cite{Herzog:2010vz,Natsuume:2018yrg}. In other words, this is the case where the scalar mass saturates the BF bound. We briefly discuss the solution for completeness. 

\subsection{Low-temperature background}

%
%

Consider the solution of the form
\begin{align}
\Psi=\Psi(u)~, A_t=A_t(u)~, A_i=0~.
%
\end{align}
Near the critical point, the scalar field remains small, and one can expand matter fields. Namely, we construct the low-temperature background perturbatively:
\begin{subequations}
\begin{align}
\Psi(u) &= \epsilon\Psi^{(1)}+ \epsilon^3\Psi^{(3)} +\cdots~, \\
A_t(u) &= A_t^{(0)}+\epsilon^2 A_t^{(2)}+\cdots~.
%
\end{align}
\end{subequations}
At zeroth order, 
\begin{align}
A_t^{(0)}=\mu(1-u^2)~.
%
\end{align}
At first order, there exists a simple solution:
\begin{align}
\Psi^{(1)} = \frac{u^2}{1+u^2}~, \quad\text{at}\quad \mu_c=\Delta=2~.
%
\end{align}
This is the solution only at the critical point. Also, this is the solution of the linear equation of motion, so the overall constant $\epsilon$ is undetermined. To resolve these issues, one needs to proceed to higher orders.

We impose the following boundary conditions:
\begin{enumerate}
\item $\Psi^{(n)}$: no fast falloff ($n\geq3$) and no slow falloff. The former means that $\calO$ comes only from $\Psi^{(1)}$. The latter is the condition for a spontaneous condensate. At the horizon, we impose the regularity condition.  
\item $A_t^{(n)}$: $A_t^{(n)}=0$ at the horizon.
\end{enumerate}
Namely, we fix the fast falloff $\calO$, but the chemical potential is corrected as
\begin{align}
\mu = \mu_c + \epsilon^2 \delta\mu_2 +\cdots~.
%
\end{align}
At higher orders,
\begin{subequations}
\begin{align}
A_t^{(2)} &= (1-u^2)\delta\mu_2 - \frac{u^2(1-u^2)}{4(1+u^2)} \\
&\sim \delta\mu_2 + \frac{1}{4}(-1-4\delta\mu_2) u^2 +\cdots~,
%
\end{align}
\end{subequations}
\begin{subequations}
\begin{align}
\Psi^{(3)} &= \frac{-2u^4+u^2(1+u^2)\ln(1+u^2)}{24(1+u^2)^2}~, \\
\delta\mu_2 &= \frac{1}{24}~.
%
\end{align}
\end{subequations}
Here, $\delta\mu_2$ is determined from the boundary condition. Then, at $O(\epsilon^2)$, the chemical potential becomes
\begin{subequations}
\begin{align}
\mu &= A_t|_{u=0} = 2+\frac{1}{24}\epsilon^2+\cdots~,\\
\Psi &\sim \epsilon u^2~, \quad(u\to0)~.
%
\end{align}
\end{subequations}
This fixes the overall constant of the condensate $\epsilon$ as
\begin{align}
\epsilon^2=24(\mu-\mu_c)~.
%
\end{align}

\subsection{Correlation length}

At high temperatures, the background solution is given by \eq{sol_high}.
%
%
Consider the linear perturbation from the background $\Psi= 0+\delta\Psi$. 
We consider the perturbation of the form $e^{-i\omega t+iqx}$. When $\Psi=0$, $\delta A_t$ and $\delta A_i$ decouple from the $\delta\Psi$-equation, and it is enough to consider the $\delta\Psi$-equation. We impose the boundary conditions (1) regular at the horizon and (2) no slow falloff. Namely, we obtain quasinormal modes. Set $\epsilon_\mu=\mu-\mu_c<0$ and employ the $(\epsilon_\mu,q)$-expansion:
\begin{align}
\delta\Psi = \psi_c + \epsilon_\mu \psi_\epsilon + q^2 \psi_q + \cdots~.
%
\end{align}
%
%
The solution is given by
\begin{subequations}
\begin{align}
\delta\Psi &\propto \frac{u^2}{4(1+u^2)} \left( -q^2 \ln u+4-2\epsilon_\mu\ln \frac{1+u^2}{u} \right) \\
&\sim -\frac{1}{4} (q^2-2\epsilon_\mu)u^2\ln u+u^2~.
%
\end{align}
\end{subequations}
Thus,
\begin{align}
q^2-2\epsilon_\mu =q^2+\xi^{-2}=0 \to \xi^2=\frac{-1}{2\epsilon_\mu}~.
%
\end{align}

\end{document}